\documentclass[aps,prb,twocolumn,superscriptaddress]{revtex4-2}

\usepackage[english]{babel}
\usepackage[utf8]{inputenc}
\usepackage{tikz}
\usetikzlibrary{calc,decorations.markings}
\usepackage{bbold}
\usepackage{amsmath}
\usepackage{physics}

\newcommand{\valos}{\mathbb{R}}
\newcommand{\eps}{\varepsilon}
\newcommand{\ugate}{\mathbb{u}}
\newcommand{\NN}{\mathcal{N}}
\newcommand{\varhatoertek}[1]{\left\langle #1 \right\rangle}
\newcommand{\VV}{\mathcal{V}}
\newcommand{\DD}{\mathcal{D}}

\newcommand{\secc}[1]{\section{#1}}

\usepackage{ifpdf}

\ifpdf
\usepackage{epstopdf}
\usepackage[pdftex,colorlinks,urlcolor=blue,citecolor=blue,linkcolor=blue]{hyperref}
\else
\usepackage[hypertex,colorlinks,urlcolor=blue,citecolor=blue,linkcolor=blue]{hyperref}
\fi
\newcommand{\elte}{\affiliation{MTA-ELTE “Momentum” Integrable Quantum Dynamics Research Group,\\
    ELTE E\"otv\"os Lor\'and University, Budapest, Hungary}}
\newcommand{\wigner}{\affiliation{ Holographic Quantum Field Theory Research Group,\\
    HUN-REN Wigner Research Centre for Physics, Budapest, Hungary}}

\begin{document}

\title{
Exact real time dynamics with free fermions in disguise
}
\author{Istv\'an Vona}
\elte \wigner 
\author{M\'arton Mesty\'an}
\elte
\author{Bal\'azs Pozsgay}
\elte

\begin{abstract}
  We consider quantum spin chains with a hidden free fermionic structure, distinct from the Jordan-Wigner transformation
  and its generalizations. We express selected local operators with the hidden fermions. This way we can exactly solve
  the real time dynamics in various physical scenarios, including the computation of selected dynamical two point
  functions, in continuous or discrete time. In the latter case we build a quantum circuit that
  can be implemented on a quantum computer. With this we extend the family of classically simulable quantum many-body
  processes. 
\end{abstract}

\maketitle

\secc{Introduction}
Free fermions play an important role in multiple areas of theoretical physics, due to their exact solvability and the
simplicity of the computations with them.
The two-dimensional classical Ising
model is solvable by free fermions \cite{Schulz-Mattis-Lieb}, just like many one dimensional quantum spin chains, such
as the Ising chain or the XX model \cite{XX-original}.
The free fermionic Kitaev chain \cite{kitaev-chain} is an important candidate for fault tolerant quantum computing.
In two dimensions the honeycomb lattice model is a free fermionic system that hosts 
anyons \cite{kitaev-honeycomb}.
Free fermions are important also in quantum information theory: quantum circuits based on
the so-called matchgates are free fermionic, and they enable classical simulability of quantum processes
\cite{matchgates1,matchgates2,matchgates3}. Free fermions also appear in the tensor network models of the holographic
principle, where they lead to an efficient contraction of the network \cite{holocode-review-jahn-eisert}.

Given their importance it is natural to ask: What is the widest class of physically meaningful models which can be
solved by free fermions? How can we construct free fermionic operators in systems that are inherently bosonic? And what is
the true computational advantage of the free fermions?

In quantum spin chains fermions can be constructed using the Jordan-Wigner (JW) transformation  \cite{jordan-wigner} (for
generalizations see \cite{japan-JW-gen-1,japan-JW-gen-2,japan-JW-gen-3,japan-free-fermion-JW,chapman-jw, Wang:2023pkp}). However, the
JW transformation does not encompass all possibilities. 
In the last couple of years a number of models have been found which can be solved by 
hidden free fermionic structures
\cite{ffd,alcaraz-medium-fermion-1,alcaraz-medium-fermion-2,fermions-behind-the-disguise,unified-graph-th,sajat-FP-model}
(for earlier examples see \cite{cooper-anyon,gyuri-susy-1,gyuri-susy-2}).
These models typically involve 4-body or higher interactions when expressed using the 
JW fermions; but the Hamiltonian can be diagonalized by the hidden fermionic operators. These are
highly non-local in the original spin operators, but also in the JW fermions. {The solution of hidden-fermion models is also different
from a generalized JW transformation in that it is not possible to express every local term in the Hamiltonian
individually as a bilinear
in the hidden fermions \cite{japan-free-fermion-JW,chapman-jw,fermions-behind-the-disguise}.} Previous works focused on computing the spectrum and the ground state properties of such models
\cite{alcaraz-medium-fermion-3,rodrigo-ffd-powerful,rodrigo-ising-and-ffd,rodrigo-random-ffd,free-fermion-subsystem-codes}
but the practical 
advantage of the free fermionic structures has not yet been demonstrated beyond the computation of the spectrum.

In this work we establish contact with local physics, focusing on the ``free fermions in
disguise'' (FFD) model of Fendley \cite{ffd}.
{For the first time we express selected local operators using the hidden fermions,
thereby partially solving the ``inverse problem''.
Afterwards we treat the real time  dynamics analytically in various scenarios. Our work is the first one to compute
such analytic results, and with this
we  demonstrate the practical usefulness of  hidden free fermions. }
 
Besides Hamiltonian dynamics we also consider discrete time evolution in quantum circuits with special geometries, compatible
with the FFD model. These circuits can 
be realized in present-day quantum 
computers. With this we demonstrate that free fermions in disguise lead to
classically simulable quantum processes and this could be used for benchmarking quantum computers.

\secc{Model} We consider one dimensional quantum spin chains and quantum circuits which are built on selected
representations of an abstract algebra. We call it the FFD algebra.
It is defined by the
generators $h_j$, $j=1, 2, \dots, M$ which satisfy 
$h_j^2=1$, $h_j=h_j^\dagger$ and the commutation relations
\begin{equation}
  \begin{split}
    \{h_j,h_{j+1}\}=\{h_j,h_{j+2}\}=0\\
    [h_j,h_k]=0,\quad |j-k|>2\,.
  \end{split}
  \label{eq:algebra}
\end{equation}
These relations are understood without periodicity in the indices, therefore the algebra always describes a system
with open boundaries. 

We consider a family of models defined by the Hamiltonians
\begin{equation}
  \label{Hdef}
  H=\sum_{j=1}^M b_j h_j
\end{equation}
where $b_j\in\valos$ are arbitrary coupling constants.
We specify a
representation of the abstract algebra on a spin 1/2-chain of length $L=M$. The concrete operators are given by
$h_1=X_1$, $h_2=Z_1X_2$, and 
\begin{equation}
\label{hrep1}
h_j=Z_{j-2}Z_{j-1}X_{j},\quad j\ge 3\,.
\end{equation}
Here and below $X_j$, $Y_j$ and $Z_j$ stand for the Pauli matrices acting on site $j$ of the spin chain. Other
representations were treated in \cite{ffd,sajat-floquet}.
{The results to be presented below do not depend on the concrete representation: different representations only change
certain degeneracies, but the techniques we develop are not affected by this.}

\secc{Jordan-Wigner solvability}

Introducing standard Jordan-Wigner fermions as 
\begin{align}
	\chi_{2k-1}&=Z_k\prod_{j=1}^{k-1} X_j, & \chi_{2k} &=Y_k \prod_{j=1}^{k-1} X_j
\end{align}
we see that the $h_j$ are quartic in the fermions
\begin{equation}
	h_{j}=-\chi_{2j-4}\chi_{2j-3}\chi_{2j-1}\chi_{2j}, 
\end{equation}
for $j\ge 3$. One might wonder whether the model is solvable by a generalized Jordan-Wigner transformation. It was proven in
\cite{fermions-behind-the-disguise} that this is not possible: there are no Majorana fermions $\tilde\chi_a$ such 
that every operator $h_j$ is expressed as a product of two Majorana fermions \cite{japan-free-fermion-JW,chapman-jw} like
\begin{equation}
	h_j = i \tilde{\chi}_{a_j} \tilde{\chi}_{b_j}, \quad \forall j,
\end{equation}
for a given mapping from $j$ to the pairs of indices $a_j,b_j$.

However, the parameters of the model can be specified so that it becomes Jordan-Wigner solvable.
For example, if we set every third $b_j$ coupling to zero, then we can introduce a re-labeling for the remaining terms as
\begin{align}
	b_{3j}&=0: & \tilde h_{2k-1} &=h_{3k-2}, & \tilde h_{2k} &= h_{3k-1}, 
\end{align}	
for $j=1, 2, \dots, \left\lfloor M/3\right\rfloor$ and $k=1,2,\dots, \left\lfloor M/3+1/2 \right\rfloor$. In a similar way we can switch off every second coupling and define the new set of Hamiltonian densities as
\begin{align}
	b_{2j}&=0: & \tilde{h}_{k} &=h_{2 k - 1}, 
\end{align}	
with $j=1,2,\dots,\left\lfloor M/2\right\rfloor$ and $k=1,2,\dots,\left\lceil M/2\right\rceil$. We find that in both cases these operators satisfy the reduced algebra
\begin{equation}
  \{\tilde h_{j},\tilde h_{j+1}\}=0,\qquad   [\tilde h_{j},\tilde h_{k}]=0 \; \text{ for } \;  |j-k|>1.
\end{equation}
These relations describe the terms in the Hamiltonian in the quantum Ising chain, and the model becomes
Jordan-Wigner solvable \cite{free-parafermion,japan-JW-gen-1,chapman-jw,japan-free-fermion-JW}. In this special case
all our results below become identical to those obtained by the standard methods; detailed comparisons will be presented
elsewhere. In the following we consider generic choices of the parameters $b_j$.

\secc{Solution}
The model possesses a family of extensive conserved charges. They are given by the logarithmic derivatives of a
transfer matrix $T_M(u)$, which can be defined via the recursion relation
\begin{equation}
  \label{Trecurs}
  T_M(u)=T_{M-1}(u)-ub_Mh_MT_{M-3}(u)\,,
\end{equation}
with the initial conditions $T_M(u)=1$ for $M\le 0$. The order of $T_M(u)$ in $u$ is $S=[(M+2)/3]$.
The Hamiltonian is obtained as $H=-\partial_u T_M(u)|_{u=0}$.  The
transfer matrices satisfy $[T_M(u),T_M(v)]=0$, therefore they also commute with the Hamiltonian. Furthermore, 
they satisfy an inversion relation
\begin{equation}\label{eq:inversion}
 T_M(u)T_M(-u)=P_M(u^2)\cdot \mathbb{1}.
\end{equation}
 Here $P_M(x)$ is a polynomial of order $S$, which is given by the recursion relation
 \begin{equation}
  \label{Precursion}
  P_M(x)=P_{M-1}(x) - x b_M^2 P_{M-3}(x)
 \end{equation}
where $P_{-2}(x) \equiv P_{-1}(x)\equiv P_0(x) \equiv -1 $.
 These polynomials and their generalizations were studied also in the
mathematical literature 
\cite{polynomial-recursions-for-ffd,polynomial-recursions-for-ffd-2}; the values $P_M(-1)$ for $b_j\equiv 1$ are known as
Narayana's cows sequence \cite{narayana-1,narayana-oeis}.

It was proven in \cite{ffd} that the model possesses a hidden free fermionic structure. There exist operators $\Psi_k$ and $\Psi_{-k}\equiv \Psi_k^\dagger$, $k=1, 2, \dots S$, 
such that 
\begin{align}
	\{\Psi_k,\Psi_\ell\}&=\{\Psi_{-k},\Psi_{-\ell}\}=0, & \{\Psi_k,\Psi_{-\ell}\}&=\delta_{k,\ell},
\end{align}
which diagonalize the Hamiltonian via
\begin{equation}
  \label{Hexpress}
  H=\sum_{k=1}^S   \eps_k [\Psi_k,\Psi_{-k}].
\end{equation}
Here the $\eps_k$ are interpreted as (half of the) single particle energies associated with the eigenmodes. They are
given by $\eps_k=(u_k)^{-1}$, with $u_k$ being the roots of a polynomial $P_M(u^2)$.

It follows from \eqref{Hexpress} that the energy levels of the Hamiltonian are given by
\begin{equation}
  E=\sum_{k=1}^S \pm \eps_k.
\end{equation}

There are only $S=[(M+2)/3]$ fermionic eigenmodes for a size $M$ of the algebra, and this is the same number  for all
faithful representations.
This implies that
each level is degenerate with the same level of degeneracy which increases exponentially with
$M$. The size of the degeneracies depends on the representation, because the length $L$ of the spin chain for a given
$M$ can vary. In our case $L=M$, therefore the the size of the degenerate sectors is $2^{M-S}$. For other representations
see
\cite{ffd,sajat-floquet}.

The physical properties of the model depend on the choice of the coupling constants $b_j$. It is natural to choose 
staggered couplings with period 3:
\begin{equation}
  b_{3j}=\alpha, \quad b_{3j+1}=\beta,\quad b_{3j+2}=\gamma.
\end{equation}
The resulting phase diagram was studied in \cite{ffd}. The set of the energy levels has a gap for generic $\alpha,
\beta, \gamma$, but there are gapless lines in the phase diagram, and the homogeneous choice $\alpha=\beta=\gamma$
corresponds to a special multi-critical point. We refer to  \cite{ffd} for more details.

\secc{Fermions} The construction of the fermionic operators is based on the so-called edge operator $\chi_0$. This is a
Hermitian operator satisfying $\chi_0^2=1$,  $\{\chi_0,h_1\}=0$ and $[\chi_0,h_j]=0$ for $j>1$. 
For certain values of $M$ we can choose $\chi_0$ to be a member of the FFD algebra \cite{sajat-floquet}, but it can also be
an extra element. In the representation \eqref{hrep1} we can choose $\chi_0=Z_1$. 

The explicit formula for the fermionic operators is \cite{ffd}
\begin{equation}
  \label{Psidef}
  \Psi_{\pm k}=\frac{1}{\NN_k}T_M(\mp u_k) \chi_{0}T_M(\pm u_k),
\end{equation}
where $\NN_k$ is a known normalization factor, see \eqref{eq:normfactor}. 
The $\Psi_k$ depend on the choice of the auxiliary element $\chi_0$; this reflects a gauge
freedom arising from the degeneracies \cite{ffd,sajat-floquet}. However, correlation functions of the elements of the
FFD algebra are independent of this choice.

\secc{\label{sec:inv}The inverse problem} In order to compute correlation functions one needs to express local operators using the
hidden fermions. In Jordan-Wigner solvable models with periodic boundary conditions the fermionic eigenmodes are Fourier
transforms 
of local fermionic operators, and the inverse problem is solved by the inverse Fourier transform. However, the situation
is more complicated in the FFD algebra, because \eqref{Psidef} is more involved than a Fourier transform. 
Furthermore, the FFD algebra has only $S$  fermionic eigenmodes for a given size $M$, therefore the inverse
problem can not be fully solved, and we can only expect solutions for selected local operators.

Our strategy is to expand $\chi_{0}$ as a linear combination of the fermionic operators as
\begin{equation}
  \label{chiansatz}
 \chi_{0}=\sum_{j=-S,j\neq0}^{S}C_{j} {\Psi}_{j}, 
\end{equation}
and to compute the $C_j$.
Afterwards we construct a family of operators which are bilinears in the
fermions. Using the results of \cite{ffd} we show in Appendix \ref{sec:decomp} that the co-efficients are
\begin{equation}\label{eq:Ck}
	C_{j} = C_{-j} = \sqrt{\frac{P_{M\setminus 1}(u_j^2)}{-u_j^2 P_{M}'(u_j^2)}} \,, 
\end{equation}
where $P_{M\setminus 1}(x)$ stands for a polynomial which is obtained analogous to $P_M(x)$ but with
the substitution $b_1=0$ and the prime in $P_M'$
denotes differentiation with respect to its argument. 

It remains to be checked whether the expansion above is  complete. We performed numerical tests for
small values of $M$ and discovered that for $M=3k$ and $M=3k+2$ the model possesses a Majorana zero mode as well. This
is described by a Hermitian operator $\Psi_0$ with the properties
$(\Psi_0)^2=\mathbb{1}$, $[H,\Psi_0]=0$ and $\{\Psi_0,\Psi_k\}=\{\Psi_0,\Psi_{-k}\}=0$ for $k=1, 2, \dots S$. The zero
mode is not a boundary mode: it is de-localized as the other fermionic modes - see \eqref{eq:decomp0} and \eqref{eq:Qu}.

If we add the zero mode as 
\begin{equation}
\label{chiansatz0}
  \chi_{0}=\sum_{j=-S}^{S}C_{j} {\Psi}_{j},\quad C_0 = 
	\begin{cases} 	0 & M\in3\mathbb{N}+1\,,\\
			\prod_{k=1}^{S}\frac{u_{k}}{\hat{u}_{k}} & \text{else},
	\end{cases} 
\end{equation}
where $\hat{u}_k > 0$ are the roots $P_{M\setminus 1}(\hat{u}_k^2)=0$, then numerical tests and an analytic proof
show in Appendix \ref{sec:decomp} that the expansion is complete, and we solved the inverse problem of $\chi_0$.

We now express other operators using the fermions, by employing the FFD algebra together with \eqref{Hexpress}. A
simple way is to construct a series of operators $o_j$ via the recursion 
\begin{equation}
	o_j=\frac{1}{2} [H,o_{j-1}],
\end{equation}
together with the initial condition $o_0=\chi_0$. This can be seen as a Krylov basis in operator space. {Such an approach was already used
in \cite{unified-graph-th} to establish the existence of hidden free fermions in related models. Here we use it simply
to find the first few examples of composite local operators, which take a simple expression in terms of the fermionic eigenmodes.}

$H$ is bilinear in the fermions,
while $o_0=\chi_0$ is linear, therefore every $o_j$ is also linear in the fermions. In the FFD algebra representation
every $o_j$ includes a factor of $\chi_0$. Having found the $o_j$, {we construct the products $o_jo_k$ and
take their linear combinations, to obtain a family of local operators that are bi-linear in the fermions. It is crucial that
$\chi_0$ drops out from every such product.}

For the first element we find $o_1=b_1 h_1\chi_0$. Taking the product $ o_1 \chi_0 = b_1 h_1$ we compute $h_1$ as a bi-linear
expression in the fermions. It follows from \eqref{Hexpress}, \eqref{chiansatz0}, and the fermionic algebra that 
\begin{equation}
  \label{h1}
	h_1= b_1^{-1} \sum_{j,k=-S}^S \eps_j C_jC_k \Psi_j \Psi_k\,.
\end{equation}
Going further we find $o_2=((b_2 h_2+b_3 h_3)b_1 h_1+b_1^2)\chi_0$, and using this result we can 
express the combination $b_2 h_2+ b_3 h_3$ as a bi-linear in the fermions. However, we did not find a way to express $h_2$ and/or
$h_3$ individually using the fermions. Afterwards the next simplest operator for which a bi-linear expression is found
is a linear combination of $h_4$,  $h_5$ and $h_2 h_3 h_5$ as discussed in Appendix \ref{sec:krylov}. Our method expresses increasingly complicated local 
operators using the fermions. At present we do not have a full classification as to which local operator is bi-linear, or perhaps of higher order in the hidden fermions.

\secc{Dynamics}
We compute infinite temperature two-point functions of operators localized around the boundary.
Our main example is the self-correlator of the boundary energy density 
\begin{equation}
  D(t)= \varhatoertek{h_1(t)h_1(0)}\equiv \text{Tr}(h_1 e^{-iHt}h_1 e^{iHt})/\text{Tr}(\mathbb{1})\,.
\end{equation}
This correlation function can be expressed as
\begin{equation}
  \label{DB}
	D(t)=\frac{1}{4 b_1^2} \left((\dot B(t))^2 - B(t)\ddot B(t) \right),
\end{equation}
where the dot means the time-derivative and the function $B(t)$ is the correlator of the edge operator itself
\begin{equation}
	B(t)= \varhatoertek{\chi_0(t)\chi_0(0)}.
\end{equation}
As explained in Appendix \ref{sec:diff} formula \eqref{DB} is the consequence of the fact that the time-evolved Krylov-basis elements $o_j(t)$ can be generated by repeatedly differentiating $\chi_{0}(t)$ w.r.t $t$,
and also Wick's theorem and the time-translation invariance of the correlators.

The time evolution of the fermionic modes is given by $\Psi_{\pm k}(t)=e^{\pm 2i\eps_{k} t} \Psi_{\pm k}$, and a
standard computation yields
\begin{align}
	B(t)&=\sum_{k=0}^S C_k^2\cos(\theta_k), & \theta_{k}&\equiv 2\eps_k t
\end{align}	
while substituting it into \eqref{DB} gives
\begin{equation}
D(t)=
		\sum_{k,\ell=0}^{S} \frac{C_{k}^{2}C_{\ell}^{2}}{4 b_1^2}  \sum_{\sigma=\pm}(\eps_{k}-\sigma\eps_{\ell})^{2}
	    \cos(\theta_{k}+\sigma \theta_{\ell}).
\label{h1corr}
\end{equation}

Note that a typical free-fermionic calculation starting out directly from formula \eqref{h1} leads to the same result. We introduced the extra step and $B(t)$ as the latter is a good building block for obtaining correlators of the more complicated local operators discussed in Section \ref{sec:inv}, and it was also simpler to analyze this function in the thermodynamical limit as explained in Section \ref{sec:res}.

\secc{\label{sec:res}Results} The above formula can be evaluated in polynomial time for every $M$ and arbitrary set of coupling
constants. It is natural to consider parameters with a period 3 staggering, in which case the model displays a rich
phase diagram  \cite{ffd}. We numerically evaluated the correlation function $D(t)$ for different values of $M$ and
various choices of the staggered parameters. We found that the correlation function can show various
types of behaviour: it can decay to zero with some power law, it can converge to a non-zero value, and it can also show
persistent oscillations around a non-zero value. Similar variation of dynamical boundary correlations was observed
earlier in standard free fermionic systems
\cite{early-boundary-autocorr,boundary-vs-bulk-autocorr,ising-boundary-edge-modes}; a full analysis of the behaviour of
$D(t)$ will be presented elsewhere.

Here we focus on the completely homogeneous case with $b_k=1$ for $k=1,\dots,M$. This is a multi-critical point in the
phase diagram, where
the energy gap (determined by the smallest $\eps_k$) scales as $M^{-z}$ with the unusual exponent
$z=3/2$. We computed a closed form result for $B(t)$ in the $M\to\infty$ limit,
by transforming the finite sum into a contour integral. Our final result reads
\begin{equation}
  B(t)=\int_{0}^{\pi}\mathrm{d}p\ C^{2}(p)\cos(2t\eps(p)),
\end{equation}
where $p$ is a momentum-like variable, and $\eps(p)$ and $C(p)$ are known functions, defined in \eqref{eq:disprel} and \eqref{eq:Cp} respectively. We also
showed that $B(t)$ can be expressed alternatively using the generalized hypergeometric function $\,_{p}F_{q}(a;b;z)$ as
\begin{equation}\label{eq:hypergeom}
	B(t) = \,_{2}F_{3}\left(\frac{1}{3},\frac{2}{3};\frac{1}{2},1,\frac{3}{2}; -\frac{27 t^{2}}{4}\right),
\end{equation}
see also \eqref{eq:sumform}. The correlation function $D(t)$ is computed afterwards via \eqref{DB}.

The long-time
asymptotics of $D(t)$ looks as
\begin{equation}\label{eq:D13p6}
  D(t)\approx \alpha
\frac{ \sin\left(3\sqrt{3}t+3\pi/4\right)}{t^{13/6}}\,,
\end{equation}
where $\alpha$ is a known constant from \eqref{eq:Dcombined}. The unusual scaling exponent $13/6$ arises from the combination of two
asymptotic contributions to $B(t)$,
see \eqref{eq:Bcombined}. 

\begin{figure}
  \centering
  \includegraphics{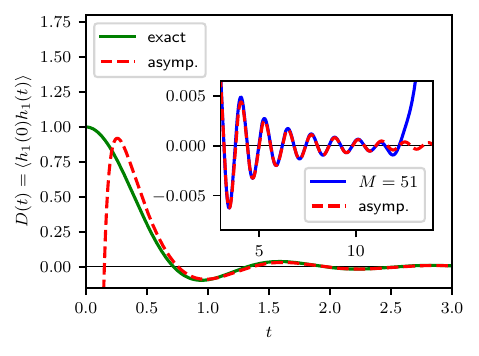}
  \caption{The boundary correlation function $D(t)$ in the thermodynamic limit $M\to\infty$, plotted against the
    asymptotic formula \eqref{eq:D13p6}. The inset shows a comparison for finite $M$.
}
  \label{fig:corr}
\end{figure}

In Figure \ref{fig:corr} we plotted the function $D(t)$ for finite $M$ and also in the $M\to\infty$ limit, together with
the asymptotic form given above. More details about the numerical computations are presented in Appendix \ref{sec:comp}.

\secc{Quantum circuit} Let us now return to finite $M$ and consider discrete time evolution in a quantum circuit. Our
aim is to derive a circuit that can be realized in present-day quantum computers, and which is compatible with the
hidden free fermions \cite{sajat-floquet}. 

The transfer matrix $T_M(u)$ admits a factorization, which yields a unitary
quantum circuit if $u$ is purely imaginary \cite{ffd,sajat-floquet}. This way we obtain a unitary matrix $\VV(\delta)$
parametrized by $\delta\in\valos$, given by
\begin{equation}
  \label{V1}
	\VV(\delta)=G\cdot G^T,\quad   G=\ugate_1 \ugate_2\cdots \ugate_{M-1} \ugate_M,
\end{equation}
where  $\ugate_j=e^{i\varphi_j h_j}$ are local unitary gates with the angles $\varphi_j$ given by the recursion
\begin{equation}
  \label{rec3}
 \tan(2\varphi_{j})=- \delta b_{j}\cos(2\varphi_{j-1})\cos(2\varphi_{j-2})
\end{equation}
together with $\varphi_0=\varphi_{-1}=0$. For small values of $\delta$ we get $\VV(\delta)\approx 1-i \delta H$,
therefore $\VV(\delta)$ can be seen as a 
``free fermionic Trotterization'' of the FFD Hamiltonian, with $\delta$ playing the role of the discretization time step. Note that the circuit is free fermionic for any choice of the $\varphi_j$-s, as the $b_j$ parameters can be chosen arbitrarily.

We choose the circuit $\VV(\delta)$ to generate discrete time evolution.  
We define dynamical two-point functions as
\begin{equation}
  \label{twop2}
 \mathcal{D}(\delta,N)=\text{Tr}(h_1 \VV(\delta)^N h_1 \VV(-\delta)^{N})/\text{Tr}(\mathbb{1})\,.
\end{equation}
In the original work \cite{ffd} the factorization of the form \eqref{V1} was derived for the transfer matrix, but it is
the novelty of this work and the parallel work \cite{sajat-floquet} to use this factorization to build unitary circuits
with local gates. We use the formulas of \cite{ffd}  specialized to the unitary case, we extend them with our solution of the
inverse problem, and this way we compute the answer for $\DD(\delta,N)$.
We obtain formally the same formula as in \eqref{h1corr}, but now with the phases $\theta_k=2 N\atan(\eps_k
\delta)$, for details see Appendix \ref{sec:floquet}.

\begin{figure}[t]
	\centering 
	\includegraphics[width=0.45\textwidth]{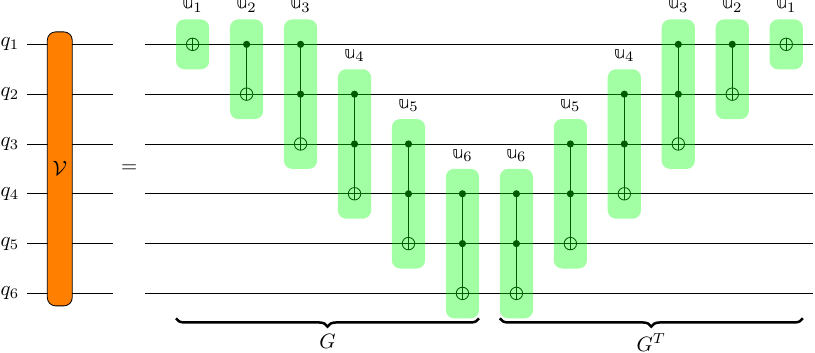}
	\caption{The staircase-like structure of $\VV$ for $M=6$ qubits}
	\label{fig:staircase}
\end{figure}

In the representation \eqref{hrep1} the local gates $\ugate_j$ are controlled unitaries in the $Z$-basis, and they can be decomposed into 4 CY gates and 1 single qubit rotation as depicted in Figure \ref{fig:3site}, for details see Appendix \ref{sec:factorization}
and \cite{controlled-unitary-decomp}. The complete circuit geometry of \eqref{V1} is shown in Figure \ref{fig:staircase}. Therefore 
the circuit $\VV(\delta)$ can be implemented on a quantum computer.
\begin{figure}[t]
\centering
\includegraphics[width=0.125\textwidth]{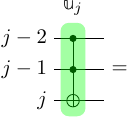}
\includegraphics[width=0.3\textwidth]{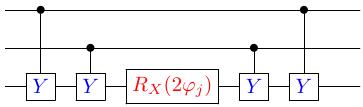}
\caption{Quantum circuit of the 3-site operator $\ugate_{j}$}
\label{fig:3site}
\end{figure}

\secc{Discussion} We used the hidden free fermionic structure to compute the time evolution of selected physical
observables. All the problems considered here can be solved in polynomial time in $M$, including finding the roots $u_k$ and
evaluating the summations for the obersvables. This means that we can classically simulate the corresponding quantum many-body processes.

Our framework should be compared to
circuits made up of matchgates \cite{matchgates1,matchgates2,matchgates3}, which are two-site gates whose
logarithm 
is bi-linear in the Jordan-Wigner 
fermions. Any product of the matchgates is still free fermionic; the only requirement is
that each gate should act on two neighbouring sites.

By contrast, in our framework the gates in the circuit need to be arranged in a special way to preserve the hidden free
fermionic property. Here we treated only one arrangement, while other circuit arrangements are
explored in  \cite{sajat-floquet}. Currently it is not known what is the most general class of circuits compatible with
the free fermions, and this is a direction for future research.

There is another limitation regarding the set of operators whose correlators can be calculated - in the efficient way outlined in the paper - on the basis of the hidden fermionic modes.
We presented examples in Section \ref{sec:inv}, but currently we do not have a full characterization of the allowed operators and their locality properties.
We hope to return to this question in future work.

In this work we only treated one specific model, namely the original model of \cite{ffd}. However, we expect that our 
techniques can be applied also to the more general class of models with hidden free fermions, treated in
\cite{fermions-behind-the-disguise,unified-graph-th} (see also \cite{sajat-FP-model}).

\begin{acknowledgments}
We are thankful to J\'anos Asb\'oth, Paul Fendley, Esperanza Lopez, Lorenzo Piroli, Roberto Ruiz, German Sierra,
Alejandro Sopena, Eric Vernier for stimulating 
discussions. We also benefitted from the workshop ``Exactly Solved Models and Quantum Computing'' organized at the
Lorentz Center, Leiden, The Netherlands, in March 2024. The authors were supported by the Hungarian National Research,
Development and Innovation Office, NKFIH Grant No. K-145904 and B.P. was supported by the NKFIH excellence grant
TKP2021\_NKTA\_64.  
\end{acknowledgments}

\appendix

\section{Basic properties of the transfer matrix}\label{sec:basic}

Here we collect a few basic properties of the transfer matrix $T_M(u)$ and the associated polynomials \cite{ffd}. In the main text
the operator $T_M(u)$ is defined via the recursion relation \eqref{Trecurs}, which can be evaluated in a straightforward
way. Alternatively, we can also express it as the power series
\begin{equation}
  T_M(u)=\sum_{\alpha=0}^S (-u)^\alpha Q^{(\alpha)},
\end{equation}
where $Q^{(\alpha)}$ is a set of commuting non-local charges, together with $Q^{(0)}=\mathbb{1}$. Introducing $H_k=b_kh_k$ the explicit
formula for $Q^{(\alpha)}$ is given by 
\begin{equation}
  Q^{(\alpha)}=\sum_{j_{k+1}>j_{k}+2}  H_{j_1}H_{j_2}\dots H_{j_\alpha}\,.
\end{equation}
Here the summation is over those subsets of indices $(j_1,j_2,\dots,j_{\alpha})$ where every pair of operators
$H_{j_k}$, $H_{j_\ell}$ commutes, leading to the condition $j_{k+1}>j_k+2$. The equivalence between the two expressions
for $T_M(u)$ can be proven by induction over $M$.

An important property of the transfer matrix is the inversion relation \eqref{eq:inversion}.
The polynomials $P_M(x)$ that arise there are defined by the recursive relations \eqref{Precursion} or alternatively via
\begin{equation}
  P_M(x)=\sum_{\alpha=0}^S  (-x)^\alpha \sum_{j_{k+1}>j_{k}+2} (b_{j_1}b_{j_2}\dots b_{j_\alpha})^2\,.
\end{equation}
For completeness we note that $P_M(x)$ can be interpreted as a weighted {\it independence polynomial} of the so-called
frustration graph of the Hamiltonian, for details see \cite{unified-graph-th}.

In the computations below we also use the restricted transfer matrix $T_{M\setminus 1}(u)$, which is obtained from
$T_M(u)$ simply by setting $b_1=0$. It can be also understood as the transfer matrix of the chain of length $M-1$ with
coupling constants $(b_2, b_3,\dots, b_M$). The associated polynomial is denoted by $P_{M\setminus 1}(u)$.

\section{\label{sec:decomp}Decomposition of the edge operator}

Here we show the edge operators' expansion via the fermionic modes
and discuss the completeness of said expression.

Let us consider the operator valued function 
\begin{equation}
\phi_{M}(u)=-\frac{1}{u}\left(\frac{P_{M}(u^{2})\chi_{0}+T_{M}(-u)\chi_{0}T_{M}(u)}{2P_{M}(u^{2})}\right)\label{eq:phiM}
\end{equation}
and its integral over the double keyhole contour depicted in Figure
\ref{fig:opcontour}. 
\begin{figure}[b]
\centering
\scalebox{1}{
\begin{tikzpicture}  

\def\gap{0.45}  
\def\bigradius{3}  
\def\littleradius{\gap/2} 
\def\xinfval{0.5}

\draw [help lines,->] (-1.25*\bigradius, 0) -- (1.25*\bigradius,0); 
\draw [help lines,->] (0, -1.25*\bigradius) -- (0, 1.25*\bigradius);

\draw[line width=1pt,   decoration={ markings,   
	mark=at position 0.2455 with {\arrow[line width=1.2pt]{<}},   
	mark=at position 0.345 with {\arrow[line width=1.2pt]{<}},	
	mark=at position 0.465 with {\arrow[line width=1.2pt]{<}},   
	mark=at position 0.7455 with {\arrow[line width=1.2pt]{<}},   
	mark=at position 0.845 with {\arrow[line width=1.2pt]{<}},   
	mark=at position 0.965 with {\arrow[line width=1.2pt]{<}}},   
	postaction={decorate}]   
	let  \n1 = {asin(\littleradius/\bigradius)}  in 
		(\n1:\bigradius) arc (\n1:180-\n1:\bigradius)
		-- (-\xinfval,\littleradius) arc (90:-90:\littleradius)
		-- (\n1:-\bigradius) arc (180+\n1:360-\n1:\bigradius)
		-- (\xinfval,-\littleradius) arc (270:90:\littleradius)   
		-- cycle;

\draw[line width=0pt, decoration = {markings, 
	mark=at position 0.05 with {\arrow[line width=1pt]{>}\arrow[line width=1pt]{<}}, 	
	mark=at position 0.15 with {\arrow[line width=1pt]{>}\arrow[line width=1pt]{<}}, 	
	mark=at position 0.4 with {\arrow[line width=1pt]{>}\arrow[line width=1pt]{<}}, 	
	mark=at position 0.7 with {\arrow[line width=1pt]{>}\arrow[line width=1pt]{<}}},  	
	postaction={decorate}] (\xinfval,0) -- (1.05*\bigradius,0) ;

\draw[line width=0pt, decoration = {markings, 
	mark=at position 0.05 with {\arrow[line width=1pt]{>}\arrow[line width=1pt]{<}}, 	
	mark=at position 0.15 with {\arrow[line width=1pt]{>}\arrow[line width=1pt]{<}}, 	
	mark=at position 0.4 with {\arrow[line width=1pt]{>}\arrow[line width=1pt]{<}}, 	
	mark=at position 0.7 with {\arrow[line width=1pt]{>}\arrow[line width=1pt]{<}}},  	
	postaction={decorate}] (-\xinfval,0) -- (-1.05*\bigradius,0) ;

\draw[line width=0pt, decoration = {markings, 
	mark=at position 0.555 with {\arrow[line width=1pt]{>}\arrow[line width=1pt]{<}}},  	
	postaction={decorate}] (-\xinfval,0) -- (\xinfval,0) ;

\node at (3.6,-0.2){$\text{Re} \, u$}; 
\node at (-0.4,3.6) {$\text{Im} \, u$}; 
\node at (-1.8,2.8) {$ C $};
\node at (\xinfval+1,-\gap-0.1) {$u_1 \,\quad \ldots \quad\, u_S$};
\node at (-\xinfval-1,-\gap-0.1) {$-u_{S} \;\;\, \ldots \;\; -u_{1}$};
\end{tikzpicture}}\caption{Contour integral to pick up the residues of the poles in the operator
$\phi_{M}(u)$ at $u=\pm u_{j},\ j=1,\ldots,S$ and the residue at
infinity. On the other hand the contour encloses the pole at $u=0$
whose residue is $\chi_{0}$ itself. }
\label{fig:opcontour}
\end{figure}
We use this contour integral as an auxiliary object to find the desired expansion of $\chi_0$ in terms of the fermionic operators.

The transfer matrix and thus the numerator of
$\phi_{M}(u)$ are both polynomials in $u.$ Therefore the singularities
of $\phi_{M}(u)$ on the complex $u$ plane consist only of the poles
coming from the zeros $u=\pm u_{j}$ of
\begin{equation}
P_{M}(u^{2})=\prod_{l=1}^{S}\left(1-u^{2}/u_{l}^{2}\right),
\end{equation}
and the simple pole at $u=0$ introduced by the $1/u$ factor. The set of zeros is known to be non-degenerate \cite{alcaraz-medium-fermion-2} and therefore the other poles are simple as well. The
closed contour then divides the complex plane into two regions. The
residue theorem inside gives simply $-\mathop{\text{Res}}_{u=0}\phi_{M}(u) = \chi_{0} $, while for the outside
region it picks up the contributions coming from the rest of the poles
\begin{align}\label{eq:fulldecomp}
\chi_{0}\overset{\text{inside}}{=}\frac{1}{2\pi i}\int_{C}\mathrm{d}u\ \phi_{M}(u) & \overset{\text{outside}}{=}\sum_{j=-S}^{S}\mathop{\text{Res}}_{u=u_{j}}\phi_{M}(u)
\end{align}
including the residue at infinity (picked up by the large circle part
of the contour). Here we adopted the shorthand notation where $u_{-j}\equiv-u_{j}$
for $j=1,2,\ldots,S$ and formally $u_{0}\equiv\infty$. When evaluated
individually, these terms are proportional to the $j\neq 0$ fermionic modes 
\begin{equation}\label{eq:decomp}
\mathop{\text{Res}}_{u=u_{j}}\phi_{M}(u)=\frac{T_{M}(-u_{j})\chi_{0}T_{M}(u_{j})}{-4u_{j}^{2}P_{M}'(u_{j}^{2})}=\frac{\mathcal{N}_{j} \Psi_{j}  }{-4u_{j}^{2}P_{M}'(u_{j}^{2})},
\end{equation}
while the residue at infinity for $j=0$ gives
\begin{equation}\label{eq:decomp0}
	\mathcal{Q}\equiv \mathop{\text{Res}}_{u=\infty}\phi_{M}(u)\overset{v\equiv1/u}{=}\mathop{\text{Res}}_{v=0}\left(-\frac{1}{v^{2}}\phi_{M}(v^{-1})\right)= \lim_{u\to\infty}\mathcal{Q}_u\,, 
\end{equation}
where we introduced
\begin{equation}\label{eq:Qu}
	\mathcal{Q}_u \equiv -u\phi_M(u) = \frac{1}{2}\left(\chi_{0}+\frac{T_{M}(-u)\chi_{0}T_{M}(u)}{P_{M}(u^{2})}\right).
\end{equation}
This gives yet another potentionally finite contribution that we need to analyze.
	In \cite{ffd} it was derived that 
\begin{equation}\label{eq:ffdanticomm}
	\{\Psi_{j},\chi_{0}\}=\frac{4P_{M\setminus 1}(u_{j}^{2})}{\mathcal{N}_{j}}\mathbb{1},
\end{equation}
together with the normalization of the fermions
\begin{equation}\label{eq:normfactor}
	\mathcal{N}_k = 4 \sqrt{ -u_{k}^{2}P_{M}'(u_{k}^{2}) P_{M\setminus 1}(u_{k}^{2}) },
\end{equation}
where $P_{M\setminus 1}(u_{k}^{2})$ is the restricted polynomial introduced above.
The coefficients on the r.h.s. of \eqref{eq:decomp} and \eqref{eq:ffdanticomm} are exactly the same for $j\neq 0$:
\begin{equation}\label{eq:Cj}
	C_{\pm j} = \frac{4P_{M\setminus 1}(u_{j}^{2})}{\mathcal{N}_{j}} = \frac{\mathcal{N}_{j}}{-4u_{j}^{2}P_{M}'(u_{j}^{2})} = \sqrt{\frac{P_{M\setminus 1}(u_j^2)}{-u_j^2 P_M'(u_j^2)}}.
\end{equation}	
If we substitute the decomposition \eqref{eq:fulldecomp} of $\chi_0$ into \eqref{eq:ffdanticomm} and use the anticommutation relations of the $\Psi_j$, then they anticommute with the object $\mathcal{Q}$ in \eqref{eq:decomp0} :
\begin{equation}
	\{\mathcal{Q} ,\Psi_{\pm j} \} = 0, \quad  j=1,2,\ldots,S.
\end{equation}
Furthermore,
\begin{equation}
	[H,\mathcal{Q}] = 0
\end{equation}
follows by the mode-expansion of the Hamiltonian \eqref{Hexpress}.  

There is one more property of $\mathcal{Q} = \lim_{u\to\infty} \mathcal{Q}_u$ we ought to show. Using the recursion relation 
\eqref{Trecurs} and the commutation/anticommutation
relations of $\chi_{0}$ with $h_{1}$ and the rest of the $h_{m}$-s
one can easily show that $\{T_{M}(u),\chi_{0}\}=2T_{M\setminus 1}(u)\chi_{0}$
which then may be used to move $\chi_{0}$ in $\mathcal{Q}_{u}$
at the end (or start) of the expression: 
\begin{equation}
\mathcal{Q}_{u}=\frac{T_{M}(-u)T_{M\setminus 1}(u)}{P_{M}(u^{2})}\chi_{0}=\chi_{0}\frac{T_{M \setminus 1}(-u)T_{M}(u)}{P_{M}(u^{2})}.\label{eq:QuTM}
\end{equation}
Now in the $u\to\infty$ limit the expressions above vanish, if the
cumulative order of the transfer matrices in the numerator is smaller
than that of $P_{M}(u^{2})$. This can only happen if the order $S'$
of $T_{M-1}(u)$ is smaller than $S$: 
\begin{equation}
S'=\left[\frac{(M-1)+2}{3}\right]<\left[\frac{M+2}{3}\right]=S,
\end{equation}
which happens only if $M \in 3\mathbb{N}+1$. Otherwise, by
multiplying the two forms of \eqref{eq:QuTM} together we have 
\begin{equation}
	\mathcal{Q}^{2}=\lim_{u\to\infty}{\mathcal{Q}_u}^2=\lim_{u\to\infty}\frac{P_{M\setminus 1}(u^{2})}{P_{M}(u^{2})}\cdot\mathbb{1},\label{eq:Qnorm}
\end{equation}
that is, $\mathcal{Q}$ squares to the identity, up to normalization.
	By looking at the properties above we can conclude that $\mathcal{Q}$ is proportional to a Majorana zero mode $\Psi_0$ 
\begin{equation}
	\mathcal{Q}=C_0 \Psi_0
\end{equation}
with energy $\eps_0=1/u_0 =0  $, that exists only if $M \notin 3 \mathbb{N}+1$. If we also choose its normalisation such that $\{\Psi_{0},\Psi_{0}\} = 2$, our
full set of fermionic anticommutation relations will look as 
\begin{equation}\label{eq:psianticomm}
	\{\Psi_{k,}\Psi_{k'}\}= 2^{\delta_{k,0}} \delta_{k+k'} \mathbb{1},
\end{equation}
with $k,k'=-S,\ldots,0,\ldots,S $.
Then, by \eqref{eq:Qnorm}, the constant $C_{0}$ can be fixed to
\begin{equation}\label{eq:C0}
C_{0}=\sqrt{\lim_{u\to\infty}\frac{P_{M \setminus 1}(u^{2})}{P_{M}(u^{2})}}=\begin{cases}
0 & M\in3\mathbb{N}+1\\
\prod_{k=1}^{S}\frac{u_{k}}{\hat{u}_{k}} & \text{else},
\end{cases}
\end{equation}
where $\hat{u}_{k}$ are the (positive) roots of the polynomial $P_{M\setminus 1}(u^{2})$. Finally, the decomposition of
the edge operators is found as
\begin{equation}
  \label{chiexp}
	\chi_0 = \sum_{j=-S}^S C_j \Psi_j,
\end{equation}	
with the coefficients defined in \eqref{eq:Cj} and \eqref{eq:C0}.

\section{Krylov basis in operator space}\label{sec:krylov}

Here we give more details about the computation of local operators that are bi-linear in the fermions. The operator
space Krylov basis is defined via
\begin{equation}
  o_{j}=\frac{1}{2}[H,o_{j-1}]\,,
\end{equation}
with the initial condition $o_0=\chi_0$. Using the expansion \eqref{chiexp} we find that
\begin{equation}
  o_j=\sum_{k=-S}^S (\eps_k)^j C_k \Psi_k,
\end{equation}
where it is understood that $\eps_{-k}\equiv -\eps_k$.

For the first few elements in the Krylov basis we find
\begin{equation}
  \begin{split}
    o_0&=\chi_0\\
    o_1&= H_1\chi_0\\
    o_2&=\left[(H_2+H_3)H_1+b_1^2\right] \chi_0\\
    o_3&=(H_4 H_2+H_4 H_3+H_5 H_3+b_1^2+b_2^2+b_3^2)H_1 \chi_0,
  \end{split}
\end{equation}
where we used the notation $H_j \equiv b_j h_j$ as in Section \ref{sec:basic}.
Taking products of the $o_j$ and their linear combinations we obtain operators that are bi-linear in the fermions. For
example we find
\begin{equation}
  (o_0-b_1^{-2}o_2)o_1=H_2+H_3\,.
\end{equation}
This will give the expansion
\begin{equation}
	H_2+H_3=\sum_{j,k=-S}^S (1-b_1^{-2}\eps_j^2) \eps_k C_jC_k \Psi_j\Psi_k\,.
\end{equation}
A further relatively simple combination is
\begin{align}
	(o_0-b_1^{-2} o_2)&(o_3-(b_1^2+b_2^2+b_3^2) o_1) \nonumber\\
	& = (b_2^2+b_3^2) H_4 + b_3^2 H_5+H_2 H_3 H_5\,.
\end{align}
This operator is expanded as
\begin{equation}
	\sum_{j,k=-S}^S (1-b_1^{-2}\eps_j^2) \eps_k(\eps_k^2-(b_1^2+b_2^2+b_3^2)) C_jC_k \Psi_j\Psi_k\,.
\end{equation}
It is clear from these computations that every operator product $o_j o_k$ is bi-linear in the fermions, and they can be
expressed alternatively as local operator products in the $h_j$. Furthermore, we could also consider quartic and higher
products in the $o_j$. We expect that certain combinations of $h_j$ can be expressed as simple
polynomials in the $o_j$, but we do not have a prescription to find the simplest such cases. 

\section{Free fermionic correlators}\label{sec:diff}

Here we briefly explain formula 
\eqref{DB}
and obtain the function $B(t)$ to calculate 
\eqref{h1corr}.  
The derivative relation can be shown simply from the fact that the
anticommutator of the edge-operator with itself 
is proportional to the identity, i.e.
\begin{equation}
\{\chi_{0}(t),\chi_{0}(t')\}=2B(t-t')\mathbb{1},
\end{equation}
at any time-separation.
This follows from the fact that we expressed it as a linear combination
of fermionic modes whose anticommutators all have the abovementioned
property. The Krylov basis $o_{j}$ may be generated via differentiation
$o_{j}(t)=(2i)^{-j}\frac{\mathrm{d}^{j}}{dt^{j}}\chi_{0}(t)$ of the
time-evolved $\chi_{0}$, which means that 
\begin{equation}
	h_{1}(t)= b_1^{-1} o_{1}(t)\chi_{0}(t)=\frac{1}{2i b_1}\dot{\chi}_{0}(t)\chi_{0}(t),
\end{equation}
where the dot denotes the time derivative. The correlator of $h_{1}$
is then the trace
\begin{equation}
D(t)=-\frac{1}{4 b_1^2}\mathrm{Tr}\left(\dot{\chi}_{0}(t)\chi_{0}(t)\dot{\chi}_{0}(0)\chi_{0}(0)\right)\,,
\end{equation}
where by Wick's theorem we have 3 different contractions among the
$\chi_{0}$-s. Each of them may be expressed via the function $B(t)$
and its derivatives due to time-translation invariance, and in the
end one arrives at the formula
\begin{equation}\label{eq:Dt}
D(t)=-\frac{1}{4 b_1^2}\left(\dot{B}^{2}(0)-\dot{B}^{2}(t)+B(t)\ddot{B}(t)\right),
\end{equation}
however as we will immediately see below, it turns out that $\dot{B}(0)=0$. 
We calculate $B(t)$ using the fermionic mode expansion of $\chi_{0}$, that leads to an even function of $t$:
\begin{align}\label{eq:BtCk}
	B(t)&=\text{Tr}\left(\chi_{0}(t)\chi_{0}(0)\right)/\text{Tr}(\mathbb{1})\\
	&=\sum_{k=-S}^{S}2^{\delta_{k,0}-1}e^{i 2\eps_{k}t}C_{k}C_{-k}= \sum_{k=0}^{S}C_{k}^{2}\cos(2\eps_{k}t),\nonumber
\end{align}
which after substitution into \eqref{eq:Dt} may be reformulated as 
\eqref{h1corr}. 
 Following the above steps,  we can compute each correlator
of the Krylov-basis elements $o_{j}(t)$ from $B(t)$ by differentiation.

\section{Thermodynamic limit in the homogeneous chain}\label{sec:TDL}

In this section we derive  $B(t)$ for $M\to\infty$ and uniform couplings $b_m=1$ and extract its large-$t$
asymptotics. The structure of the latter then leads to the scaling 
\eqref{eq:D13p6}
of $D(t)$.  

First, let us remark on the behaviour of the fermion energies in TDL. In appendix C of \cite{ffd}
the dispersion relation of the uniform model 
\begin{equation}
\eps^{2}(p)=\frac{\sin^{3}p}{\sin\left(p/3\right)\sin^{2}\left(2p/3\right)},\quad p\in[0,\pi]\label{eq:disprel}
\end{equation}
was derived (see Figure \ref{fig:behaviour}), and it was also argued in \cite{alcaraz-medium-fermion-1}
that in the large $M$ limit the solutions of $P_{M}(x_{k}=\eps^{-2}(p_{k}))=0,\; M\in3\mathbb{N}$
occupy the above interval in terms of the momentum variable $p$ in
an equidistant way $p_{k+1}-p_{k}\sim3\pi/M$. The interpretation of the variable $p$ as a lattice momentum was not
explicitly given in the previous works. Such an interpretation can be given by the explicit real
space representation of the fermionic operators; this computation will be presented elsewhere.

The formula
above gives the supremum of the energy (thus the infimum
of the roots) at $p=0$ as
\begin{equation}
\frac{3\sqrt{3}}{2}>\eps_{k}>0,\quad x_{\text{inf}}\equiv\frac{4}{27}<x_{k}<\infty.
\end{equation}
We then analyze $B(t)$ and start from finite system sizes. By looking at \eqref{eq:BtCk} and the forms of the $C_j$ coefficients in \eqref{eq:Cj} and \eqref{eq:C0} one may realize that it is possible to rewrite the finite sum as a contour integral
\begin{equation}
	B(t)=\frac{1}{2\pi i}\int_{C}\mathrm{d}z \cos(2t/\sqrt{z}) \left(\frac{P_{M\setminus 1}(z)}{-z P_{M}(z)}\right),\label{eq:contourint}
\end{equation}
with a keyhole contour $C$ surrounding the half-infinite line starting from $z=x_{\text{inf}}$ as shown on the left of Figure \ref{fig:contour}. 
\begin{figure}
{\centering
\scalebox{0.5}{
\begin{tikzpicture} 

\def\gap{0.45} 
\def\bigradius{3} 
\def\littleradius{\gap/2}
\def\xinfval{0.5}

\draw [help lines,->,line width=1.5] (-1.25*\bigradius, 0) -- (1.25*\bigradius,0);
\draw [help lines,->, line width = 1.5] (0, -1.25*\bigradius) -- (0, 1.25*\bigradius);

\draw[line width=2pt,   decoration={ markings,
  mark=at position 0.2455 with {\arrow[line width=2pt]{<}},
  mark=at position 0.665 with {\arrow[line width=2pt]{<}},
  mark=at position 0.805 with {\arrow[line width=2pt]{<}},
  mark=at position 0.965 with {\arrow[line width=2pt]{<}}},
  postaction={decorate}]
  let
     \n1 = {asin(\gap/2/\bigradius)}
  in (\n1:\bigradius) arc (\n1:360-\n1:\bigradius)
  -- (\xinfval,-\gap/2) arc (270:90:\littleradius)
  -- cycle;

\draw[line width=0pt, decoration = {markings,
	mark=at position 0.05 with {\arrow[line width=1.5pt]{>}\arrow[line width=1.5pt]{<}},
	mark=at position 0.08 with {\node[draw, shape=circle, fill = white, line width=1.5pt,minimum size=0.2cm, inner sep=0pt](){};},
	mark=at position 0.15 with {\arrow[line width=1.5pt]{>}\arrow[line width=1.5pt]{<}},
	mark=at position 0.20 with {\node[draw, shape=circle, fill = white, line width=1.5pt,minimum size=0.2cm, inner sep=0pt](){};},
	mark=at position 0.4 with {\arrow[line width=1.5pt]{>}\arrow[line width=1.5pt]{<}},
	mark=at position 0.5 with {\node[draw, shape=circle, fill = white, line width=1.5pt,minimum size=0.2cm, inner sep=0pt](){};},
	mark=at position 0.7 with {\arrow[line width=1.5pt]{>}\arrow[line width=1.5pt]{<}},
	mark=at position 0.85 with {\node[draw, shape=circle, fill = white, line width=1.5pt,minimum size=0.2cm, inner sep=0pt](){};}
}, 
	postaction={decorate}] (\xinfval,0) -- (1.05*\bigradius,0) ;

\node[draw,shape=circle, fill=black, line width=3pt,minimum size=0.2cm, inner sep=0pt] at (0,0){};

\node at (3.6,-0.4){\Large $\text{Re} \, z$};
\node at (-0.8,3.6) {\Large $\text{Im} \, z$};
\node at (-1.9,2.9) {\Large $ C $};
\node at (1.8, \gap+0.4) {\Large $ x + i 0 $};
\node at (1.8, -\gap-0.4) {\Large $ x - i 0$};
\node at (\xinfval+0.1, -\gap-0.1) {\Large $x_{\text{inf}}$};

\end{tikzpicture}
\begin{tikzpicture} 

\def\gap{0.45} 
\def\bigradius{3} 
\def\littleradius{\gap/2}
\def\xinfval{0.5}

\draw [help lines,->, line width = 1.5pt] (-1.25*\bigradius, 0) -- (1.25*\bigradius,0);
\draw [help lines,->, line width = 1.5pt] (0, -1.25*\bigradius) -- (0, 1.25*\bigradius);

\draw[line width=3pt, decoration={markings, mark=at position 0.5 with {\arrow[line width=1pt, scale=2]{>}}}, postaction={decorate} ] (\xinfval,0) -- (1.15*\bigradius,0) ;

\node[draw,shape=circle, fill=black, line width=3pt,minimum size=0.2cm, inner sep=0pt] at (0,0){};

\node at (3.6,-0.4){\Large $\text{Re} \, z$};
\node at (-0.8,3.6) {\Large $\text{Im} \, z$};
\node at (\xinfval+0.1, -\gap-0.1) {\Large $x_{\text{inf}}$};

\end{tikzpicture}}}

\caption{The contour $C$ for integration in \eqref{eq:contourint} and the
analytic structure of the integrand (left). The poles (crosses) of $P_{M}(x)$
and the zeros (small circles) of $P_{M\setminus1}(x)$ thought to ``condense'' into the branchcut (thick line on the right) of $d(z)$
starting from $x_{\text{inf}}$ in the $M\to\infty$ limit. The large
circle picks up the residue at infinity, which gives the contribution
of the zero mode. In the TDL the latter can be dropped and the rest
of the contour integrates the discontinuity. For $t>0$ the integrand
has a has an essential singularity (black dot) at $z=0$, while for
$t=0$ it is only a simple pole due to the factor $(-z)^{-1}$, with
residue one, giving us the sum rule $\sum_{k=0}^{S}C_{k}^{2}=1$.}
\label{fig:contour}
\end{figure}
The zeros $x_k$ of $P_M(z)$ in the denominator of the integrand introduce poles along this line, whose residues get picked up by the contour $C$ and they are exactly the same as the terms under the sum in \eqref{eq:BtCk}. The $C_{0}^{2}$ term in $B(t)$
gets reconstructed by the residue at infinity
\begin{align}
	&\text{Res}_{z=\infty}\left(\frac{\cos(2t/\sqrt{z})P_{M\setminus 1}(z)}{-z P_M(z)}\right) \qquad (w=1/z) \\
	&=\text{Res}_{w=0}\left(\frac{\cos(2t\sqrt{w})P_{M\setminus 1}(1/w)}{w P_M(1/w)}\right)=\lim_{z\to\infty}\frac{P_{M\setminus 1}(z)}{P_M(z)},\nonumber
\end{align}
that gets picked up by the large circle part of the contour. In fact, for the unifom case $b_m=1$, this latter contribution of the zero mode always vanishes
in the $M\to\infty$ limit 
\begin{equation}
C_{0}^{2}=\frac{3(r-1)}{M+2r}=\mathcal{O}(1/M),\quad M\in3\mathbb{N}+r,\quad r=1,2,3\label{eq:Czero}
\end{equation}
 as can be easily seen from the explicit expression of the polynomials that can be found in \cite{alcaraz-medium-fermion-1}.

Now we turn our attention to the $M\to \infty$ limit of the integral above. For our later convenience we introduce the rational function
\begin{equation}
  \label{dMz}
	d_M(z)\equiv\frac{P_{M\setminus 1}(z)}{P_{M}(z)}.
\end{equation}	
In the uniform case, where $P_{M\setminus1}(x)=P_{M-1}(x)$, we may rewrite the recursion 
\eqref{Precursion}  
in terms of this
object as 
\begin{equation}
  \label{dM}
1=d_{M}(x)-xd_{M}(x)d_{M-1}(x)d_{M-2}(x)
\end{equation}
If for a certain $x$ there exists the limit
\begin{equation}
d(x)\equiv\lim_{M\to\infty}d_{M}(x),
\end{equation}
then \eqref{dM} implies that the limiting value $d(x)$ has to satisfy 
the cubic equation
\begin{equation}
1=d(x)-xd^{3}(x).\label{eq:dcubic}
\end{equation}
It follows from the definition \eqref{dMz} and the behaviour of the polynomials that for large $M$ the function $d(z)$
will have a large 
number of zeroes and poles at the real line for $z>4/27$. This implies that $d_M(z)$ can not converge for these values
of $z$. Numerical experiments confirm this expectation. On the other hand, numerical investigations show that the
recursion \eqref{dM} is convergent for $\text{Im}(z)\ne 0$, $\text{Re}(z)>0$, and the limiting value is always given by one of the
roots of \eqref{eq:dcubic}.
More concretely, we find that
\begin{equation}
d(z)=\begin{cases}
d^{(+)}(z) & \text{Im}\,z>0\\
d^{(-)}(z) & \text{Im}\,z<0
\end{cases}
\end{equation}
where the two roots are
\begin{equation}
d^{(\pm)}(z)=e^{\pm i\frac{\pi}{3}}W^{\frac{1}{3}}(z)+\frac{e^{\mp i\frac{\pi}{3}}}{3zW^{\frac{1}{3}}(z)},
\end{equation}
and we introduced
\begin{equation}
W(z)\equiv\frac{1}{2z}\left(1+\sqrt{1-\frac{x_{\text{inf}}}{z}}\right).
\end{equation}	
When evaluated close to the positive real line as $d^{(\pm)}(x\pm i0),\ x\in\mathbb{R}^{+}$
, the behaviour of these two roots switches at $x=x_{\text{inf}}$
where the discriminant of the cubic equation \eqref{eq:dcubic} changes
sign: for $x<x_{\text{inf}}$ they both tend to the same real value,
whereas for $x>x_{\text{inf}}$ they are nothing but the pair of complex
conjugate roots, with a finite imaginary value. Therefore, $d(z)$
has a branchcut starting at $x_{\text{inf}}$, which runs
along the real line into $x=+\infty$.

Let us now return to the contour integral \eqref{eq:contourint}. Performing the integral surrounding the branchcut
we need to evaluate the discontinuity
\begin{align}
	\text{Disc}\ d(x)&=\frac{1}{2\pi i}\left(d^{(-)}(x-i0)-d^{(+)}(x+i0)\right) \nonumber \\
	&=-\frac{\sqrt{3}}{2\pi}\left(W^{\frac{1}{3}}(x)-\frac{1}{3xW^{\frac{1}{3}}(x)}\right). 
\end{align}
Due to \eqref{eq:Czero} we may drop the large circle from the contour integral, thus we find
\begin{equation}
B(t)=\int_{x_{\text{inf}}}^{\infty}\mathrm{d}x\ \frac{\cos(2t/\sqrt{x})\text{Disc}\,d(x)}{(-x)}.
\end{equation}
By the correspondance $x=\eps^{-2}$ this is essentially an integral
in terms of the inverse squared energy. By a change of variables via \eqref{eq:disprel}
it may be rewritten as a momentum integral
\begin{equation}
B(t)=\int_{0}^{\pi}\mathrm{d}p\ C^{2}(p)\cos(2t\eps(p)),\label{eq:pintegral}
\end{equation}
where the function $C^{2}(p)$ is given by
\begin{align}\label{eq:Cp}
	C^{2}(p)=-\frac{3}{2\pi}\frac{\mathrm{d}\eps^{2/3}}{\mathrm{d}p}&\Bigg\lbrace  \left(\eps(0)+\sqrt{\eps^{2}(0)-\eps^{2}(p)}\right)^{1/3}  \\
	&-\left(\eps(0)-\sqrt{\eps^{2}(0)-\eps^{2}(p)}\right)^{1/3}\Bigg\rbrace , \nonumber
\end{align}
as expressed in terms of the dispersion relation (see Figure \ref{fig:behaviour}).

We also found alternative representations of the same function $B(t)$. Without derivations we communicate that it can be
written also as the expansion
\begin{equation}\label{eq:sumform}
	B(t)=\sum_{k=0}^{\infty} \frac{(3k)!}{k!(2k+1)!(2k)!} \left(-4 t^2\right)^{k} 
\end{equation}
that is equivalent to the special function \eqref{eq:hypergeom}.
Detailed derivations of this alternative
representation, together with connections to the combinatorical properties of the recursion \eqref{dM} will be presented
elsewhere.

\section{Long time asymptotics}

\begin{figure}[t]
	\centering
	\includegraphics[width=0.4\textwidth]{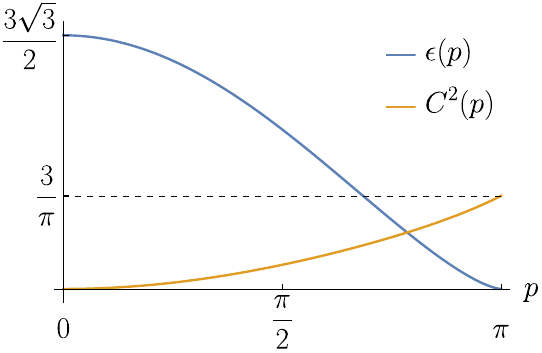}
	\caption{Behaviour of the dispersion relation $\epsilon(p)$ and the function $C^2(p)$}
	\label{fig:behaviour}
\end{figure}

The integral \eqref{eq:pintegral} is an exact representation
of \eqref{eq:hypergeom}. Here we extract its large
time asymptotics, using the stationary phase approximation. The stationary
points of the energy are exactly at the lowest and highest momentum,
and the energy and the function $C^{2}(p)$ behaves as
\begin{equation}
\eps(p)  =\begin{cases}
\eps(0)\left(1-\frac{1}{6}p^{2}+\mathcal{O}(p^{4})\right)\\
\left(\frac{4}{3}\right)^{3/4}\vert\pi-p\vert^{3/2}+\mathcal{O}\left(\vert\pi-p\vert^{5/2}\right)
\end{cases} 
\end{equation}
and
\begin{equation}
	 C^{2}(p)  =\begin{cases}
\frac{p^{2}}{3\pi}+\mathcal{O}(p^{4}) & \quad p\simeq0\\
\frac{3}{\pi}+\mathcal{O}\left(\vert\pi-p\vert\right) & \quad p\simeq\pi
\end{cases}
\end{equation}
respectively around these points, see also Figure \ref{fig:behaviour}. 
In the first case, at $p\simeq0$ we have to
evaluate (properly regulated) complex Gaussian integrals that give
\begin{align}
	\frac{1}{4}\sum_{\sigma=\pm}\int_{-\infty}^{\infty}\mathrm{d}p\ & \frac{p^{2}}{3\pi}e^{i\sigma2t\eps(0)\left(1-\frac{1}{6}p^{2}\right)} \label{eq:LOterm0} \\
	&=-\sqrt{\frac{3}{2\pi}}\frac{\sin\left(2\eps(0)t+3\pi/4\right)}{(2\eps(0)t)^{3/2}}, \nonumber
\end{align}
for $t>0$, whereas for $p\simeq\pi$, after changing variables to
$\eta=\left(\frac{4}{3}\right)^{3/4}\vert\pi-p\vert^{3/2}$ we get
\begin{align}
	& \frac{1}{2}\sum_{\sigma=\pm}\int_{-\infty}^{\pi}\mathrm{d}p\  \frac{3}{\pi}e^{i\sigma2t\left(4/3\right)^{3/4}\vert\pi-p\vert^{3/2}} \label{eq:LOtermpi}  \\ &=\frac{1}{2}\sum_{\sigma=\pm}\int_{0}^{\infty}\mathrm{d}\eta\ \frac{\sqrt{3}}{\pi}\eta^{-1/3}e^{i\sigma2t\eta}=\frac{\sqrt{3}\Gamma(\frac{2}{3})}{2\pi\times(2t)^{2/3}}. \nonumber
\end{align}
Although we could continue these approximations around the two stationary
points, and both would give a family of corrections to $B(t)$ as
inverse fractional powers of $t$ competing in magnitude, we stop
at the lowest order terms derived above
\begin{equation}\label{eq:Bcombined}
B(t)\sim\frac{B_{\pi}}{t^{2/3}}+\frac{B_{0}(t)}{t^{3/2}}.
\end{equation}
Here we indicated their structure with the constant $B_{\pi}$ and
the oscillating function $B_{0}(t)$ that can be read off from the
r.h.s. of \eqref{eq:LOtermpi} and \eqref{eq:LOterm0} respectively.
Substituting this into 
\eqref{DB}
tells us that the leading
term in the correlator of $h_{1}$ should behave with $t$ to the
inverse power with exponent $2/3+3/2=13/6$, more precisely as 
\begin{equation}\label{eq:Dcombined}
	D(t)\sim-\frac{B_{\pi}\ddot{B}_{0}(t)}{4t^{13/6}}=-\frac{3\times3^{3/4}\Gamma(\frac{2}{3})}{16\times2^{1/6}\pi^{3/2}}\frac{\sin\left(2\eps(0)t+3\pi/4\right)}{t^{13/6}},
\end{equation}
fixing the constant $\alpha\approx -0.0926$ in 
\eqref{eq:D13p6}.

\section{\label{sec:floquet}Floquet time evolution from the transfer matrix}

As it was described in the main text, the transfer matrix admits a factorization into a product of local gates, such
that for purely imaginary rapidities we obtain a unitary quantum circuit. More precisely, the transfer matrix becomes an
operator that is proportional to a unitary quantum circuit. 
In the present conventions we get
\begin{equation}
\mathcal{V}(\delta)=\frac{T_{M}(-i\delta)}{\sqrt{P_{M}(-\delta^{2})}}.
\end{equation}
Here we obtained the additional normalization factor from the inversion relation 
\eqref{eq:inversion}. 

The transfer matrix can be written \cite{ffd} 
via the $\Psi_{k\ne 0}$ operators as
\begin{equation}\label{eq:Tprod}
	T_{M}(u)=\prod_{k=1}^{S}\left(1-\frac{u}{u_k}[\Psi_{k},\Psi_{-k}]\right).
\end{equation}
Then, using solely \eqref{eq:psianticomm}, a conjugation with the transfer matrix for $k \neq 0$
\begin{equation}
\frac{T_{M}(u)\Psi_{k}T_{M}(-u)}{P_{M}(u^{2})}=\frac{u_{k}-u}{u_{k}+u}\Psi_{k}\label{eq:Tconj}
\end{equation}
will only introduce the above prefactor, where for negative $k$-s $u_{-\vert k \vert}\equiv -u_{\vert k \vert}$ is understood.
After analytic continuation $u\to-i\delta$ this turns into a phase
\begin{equation}
\frac{u_{k}+i\delta}{u_{k}-i\delta}=e^{i2\arctan(\eps_{k}\delta)}.
\end{equation}
Then, several unitaries will act as
\begin{equation}
  \mathcal{V}^{N}(\delta)\Psi_{k}\mathcal{V}^{N}(-\delta)=
  e^{i  \theta_{k}}\Psi_{k},\quad\theta_{k}= 2 N\arctan(\eps_{k}\delta).\label{eq:phases}
\end{equation}
For the $k=0$ zero mode we have $\theta_{0}=0$, as the conjugation
on the l.h.s. of \eqref{eq:Tconj} leaves $\Psi_{0}$ invariant.

At this point we have to perform very similar steps to what we
had in Section \ref{sec:TDL} to calculate the two-point function
of $h_{1}$ with itself. That is, substitute 
\eqref{h1}
into 
\eqref{twop2}
and evolve the product of two fermions
via \eqref{eq:phases}. After that, taking the trace leads to the
the same structure, the only difference is that now
one needs to compute the traces of four fermion terms directly. However, this is
straightforward and it leads to the same formula, and only the phases $\theta_k$ are modified.

\section{\label{sec:factorization}Factorization of the 3-site gate}

Here we provide an explicit factorization of the gates in the quantum circuit described in the main text.
The circuit \eqref{V1} consists of 3-site gates
\begin{equation}
  \ugate_j\equiv \exp(i \varphi_j h_j) = \exp(i \varphi_j Z_{j-2} Z_{j-1} X_{j})\,.
  \label{eq:3site}
\end{equation}
After expanding the exponential in the definition, the gate simply looks as
\begin{equation}
\ugate_{j}=\cos\varphi_{j}+i\sin\varphi_{j}Z_{j-2}Z_{j-1}X_{j}.
\label{eq:3sdef}
\end{equation}
Based on the idea of uniformly controlled unitaries in \cite{controlled-unitary-decomp} we express
this operator as a product of quantum gates. As the rotations in 
\eqref{eq:3sdef} 
are around the $x$-axis, the set of gates we use is such a 1-qubit
rotation
\begin{equation}
[R_{X}(\varphi)]_j=e^{i\frac{\varphi}{2}X_{j}}=\cos\frac{\varphi}{2}+i\sin\frac{\varphi}{2}X_{j}
\end{equation}
and controlled-Y, or CY gates:
\begin{equation}
[\text{CY}]_{k,l}=e^{i\frac{\pi}{4}\left(\mathbb{1}-Z_{k}\right)\left(\mathbb{1}-Y_{l}\right)}=\frac{1}{2}\left(\mathbb{1}+Z_{k}+Y_{l}-Z_{k}Y_{l}\right),
\end{equation}
where $k$ is the control qubit and $l$ is the controlled one. The
square of the CY gate is the identity, and inserting an $X$ operator
on the controlled qubit between two such gate gives $[\text{CY}]_{k,j}X_{j}[\text{CY}]_{k,j}=Z_{k}X_{j}$
for $k\neq j$.
This leads to the idea to represent $\ugate_{j}$ by conjugating the rotation operator
on the $j^{\text{th}}$ site twice, by two different CY gates.
In fact we obtain the factorization (see Figure \ref{fig:3site}):
\begin{equation} \label{eq:ujblock}
\ugate_{j}=[\text{CY}]_{j-2,j}[\text{CY}]_{j-1,j}[R_{X}(2\varphi_{j})]_{j}[\text{CY}]_{j-1,j}[\text{CY}]_{j-2,j}.
\end{equation}

Close to the boundary of the system, i.e. for $j=1,2$ the CY gates whose control bit would lie outside the index range $1,2,\ldots, M$ should be replaced by an identity
in formula \eqref{eq:ujblock}, as can be seen in Figure \ref{fig:staircase}.

\section{\label{sec:comp}Comparison with exact diagonalization}

\begin{figure*}
	\centering
  \includegraphics{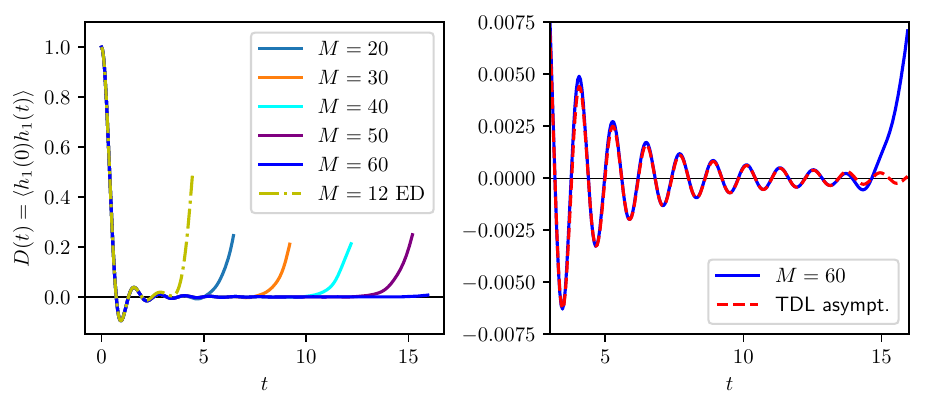}
	\caption{ Left: the dashed-dotted line shows the numerical exact diagonalization result for the energy-energy correlation function at the largest reasonable system size for an average desktop PC without parallelization~($M=12$). Full lines show the analytical result \eqref{h1corr} for the energy-energy correlation function at different system sizes $M$.  Right: comparison of the asymptotics of the analytical result at $M=60$ with the asymptotics in the thermodynamic limit \eqref{eq:D13p6}. } 
	\label{fig:prec}
\end{figure*}

To show how much more powerful the analytical method is than numerical exact diagonalization (ED) we plot the analytical results for the energy-energy correlation function \eqref{h1corr} against those given by ED (see Fig. \ref{fig:prec}). Much higher system sizes can be reached analytically than using numerical ED on a desktop PC. The difficulty that limits the evaluation of the analytical formula \eqref{h1corr} is finding all the roots of the polynomial $P_M(x)$ \eqref{Precursion}. The order of the polynomial grows as $S=[(M+2)/3]$ and thus for large $M$ it becomes difficult to find the roots. Using the \verb!NSolve! function of \verb!Mathematica! we can go up to $M=60$. Meanwhile, the largest system size reachable by numerical ED on an average desktop PC without parallelization is $M=12$.


%

\end{document}